\documentclass[superscriptaddress,groupaddress,reprint,english]{revtex4-1}
\usepackage[L7x,T1]{fontenc}
\usepackage[latin9]{inputenc}
\usepackage{xcolor}
\usepackage{pdfcolmk}
\usepackage{array}
\usepackage{bm}
\usepackage{multirow}
\usepackage{amsmath}
\usepackage{amssymb}
\usepackage{graphicx}
\usepackage{esint}
\usepackage{bm}

\makeatother

\usepackage{babel}
\begin{document}
\global\long\def\emath{\mathrm{e}}
\global\long\def\d{\mathrm{d}}
\global\long\def\vec#1{\bm{#1}}
\global\long\def\icm{\mathrm{\mathrm{cm}^{-1}}}
\global\long\def\imath{\mathrm{i}}
\global\long\def\S{\mathrm{S}}
\global\long\def\Qx{\mathrm{Q_{x}}}
\global\long\def\Qy{\mathrm{Q_{y}}}
\global\long\def\Q{\mathrm{Q}}

\title{Diversity of coherences and origin of electronic transitions of supermolecular
nanoring}

\author{Vytautas~Butkus}

\altaffiliation{These authors equally contributed to this work }

\affiliation{Department of Theoretical Physics, Faculty of Physics, Vilnius University,
Sauletekio Avenue 9-III, 10222 Vilnius, Lithuania}

\affiliation{Center for Physical Sciences and Technology, Savanoriu Avenue 231,
02300 Vilnius, Lithuania}

\author{Jan~Alster}

\altaffiliation{These authors equally contributed to this work }

\altaffiliation[Current address: ]{Faculty of Mathematics and Physics, Charles University in Prague, Ke Karlovu 3, 121 16 Prague, Czech Republic}

\affiliation{Department of Chemical Physics, Lund University, P.O. Box 124, SE-22100
Lund, Sweden}

\author{Egl\.{e}~Ba\v{s}inskait\.{e}}

\affiliation{Department of Theoretical Physics, Faculty of Physics, Vilnius University,
Sauletekio Avenue 9-III, 10222 Vilnius, Lithuania}

\author{Ram\={u}nas~Augulis}

\altaffiliation[Current address: ]{Center for Physical Sciences and Technology, Savanoriu Avenue 231, 02300 Vilnius, Lithuania}

\affiliation{Department of Chemical Physics, Lund University, P.O. Box 124, SE-22100
Lund, Sweden}

\author{Patrik~Neuhaus}

\affiliation{Department of Chemistry, University of Oxford, Chemistry Research
Laboratory, Mansfield Road, Oxford OX1 3TA, United Kingdom}

\author{Leonas~Valkunas}

\affiliation{Department of Theoretical Physics, Faculty of Physics, Vilnius University,
Sauletekio Avenue 9-III, 10222 Vilnius, Lithuania}

\affiliation{Center for Physical Sciences and Technology, Savanoriu Avenue 231,
02300 Vilnius, Lithuania}

\author{Harry~L.~Anderson}

\affiliation{Department of Chemistry, University of Oxford, Chemistry Research
Laboratory, Mansfield Road, Oxford OX1 3TA, United Kingdom}

\author{Darius~Abramavicius}

\affiliation{Department of Theoretical Physics, Faculty of Physics, Vilnius University,
Sauletekio Avenue 9-III, 10222 Vilnius, Lithuania}

\author{Donatas~Zigmantas}

\email{donatas.zigmantas@chemphys.lu.se}

\affiliation{Department of Chemical Physics, Lund University, P.O. Box 124, SE-22100
Lund, Sweden}
\begin{abstract}
Quantum coherence is highly involved in photochemical functioning
of complex molecular systems. Co-existence and intermixing of electronic
and/or vibrational coherences, while never unambiguously identified
experimentally, has been proposed to be responsible for this phenomenon.
Analysis of multidimensional spectra of a synthetic belt-shaped molecular
six-porphyrin nanoring with an inner template clearly shows a great
diversity of separable electronic, vibrational and mixed coherences
and their cooperation shaping the optical response. The results yield
clear assignment of electronic and vibronic states, estimation of
excitation transfer rates, and decoherence times. Theoretical considerations
prove that the complexity of excitation dynamics and spectral features
of the nanoring excitation spectrum is due to combined effect of cyclic
symmetry, small geometrical deformations, and vibronic coupling.
\end{abstract}
\maketitle
Electronic energy transfer (EET) in organic polymers plays a central
role in their versatile applications as electric charge carrying or
light emitting/absorbing moieties, allowing facile and cheap production
of optoelectronic devices\cite{Nalwa1997handbook}. Even though various
nanomaterials, including quantum dots or carbon nanotubes, have been
extensively studied, the interest still persists due to the great
variability of their optical and electronic properties and enormous
chemical diversity.

EET in large supra-molecular systems is necessarily related to coherent
phenomena, arising from the quantum mechanical nature of the inner
microscopic constituents. In the conditions of electronic--vibrational
resonance, quantum mixing of electronic and nuclear degrees of freedom
of vibronically coupled molecular systems were suggested to result
in a host of phenomena beyond the adiabatic approximation\cite{Jonas_PNAS2012,Butkus_vibronic_aggregatess_JCP2014}.

Recent development of ultrafast nonlinear spectroscopic techniques,
such as the two-dimensional electronic spectroscopy (2DES) facilitated
direct observation of quantum superpositions in the form of quantum
beats (dynamic coherences)\cite{engel-nat2007,Collini2009,nemeth-sperling-JCP2010,Hayes21062013,Halpin_Miller-NChem2014,Fuller_NChem_2014,Romero2014}.
Although 2DES combines correlation information between excitation
and detection frequencies and simultaneously ensures high spectral
and temporal resolutions, the experimental identification of coherences
and determination of their origin is a very complex issue. The questionable
nature of the long-lived dynamic coherences in the iconic Fenna--Matthews--Olson
(FMO) light-harvesting complex (whether electronic\cite{engel-nat2007}
or vibrational\cite{Christensson2011,Jonas_PNAS2012,Tempelaar2013})
still causes debates. Mechanisms related to electronic--vibrational
mixing\cite{Christensson_JPCB2012,Chin2013}, inter-pigment correlations\cite{abramavicius:174504},
non-secular quantum transport\cite{Panitchayangkoon2011}, and inhomogeneous
broadening\cite{Dong2014} were proposed to be important, allowing
FMO to maintain the dynamic quantum coherence for as long as a few
picoseconds.

The role of electronic--vibrational mixing in molecular systems is
of a particular interest, since it was recently shown that diabatic
coupling to coherent vibrational modes might enhance the rate of EET\cite{Womick2011,Kolli2012,Chin2013}
and of the charge transfer\cite{Fuller_NChem_2014,Romero2014}. However,
this implies existence of both electronic and vibrational coherences
in the same system at the same time, what has never been unambiguously
observed. 

\begin{figure}
\includegraphics[width=8cm]{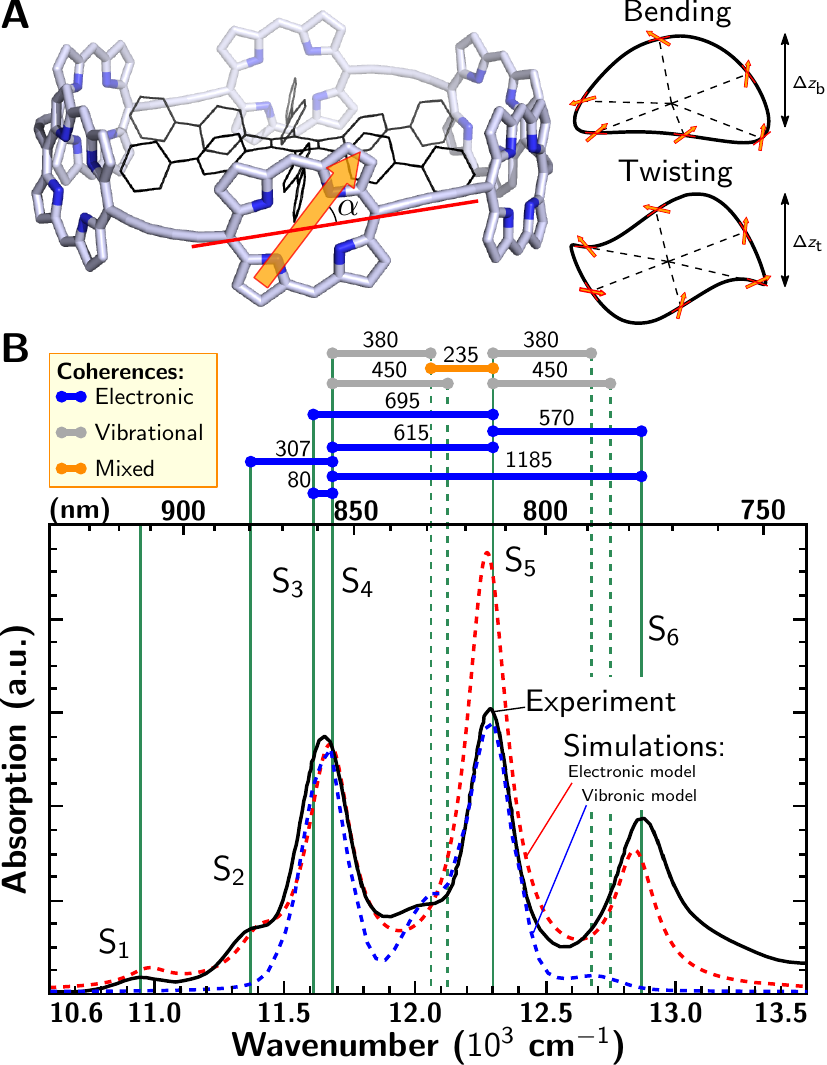}

\protect\caption{\label{fig:Structure-of-porphyrin}\textbf{Structure (A) and absorption
spectrum (B) of the porphyrin nanoring.} Deformations used in the
excitonic calculation are shown on the right. (\textbf{B}) Measured
(black solid line) and simulated (red and blue dashed lines) absorption
spectrum at 77~K. Energies of the electronic transitions to states
$\protect\S_{1}-\protect\S_{6}$ are indicated by solid vertical lines;
vibronic transitions are indicated by vertical dashed lines. Frequencies
of resolved coherent beatings are shown as colored segments, connecting
the states involved in the corresponding quantum superpositions.}
\end{figure}

Porphyrin nanoring, the chemical structure of which is depicted in
Fig.~\ref{fig:Structure-of-porphyrin}A, was chosen for this quest.
It consists of six zinc(II) porphyrin molecules forming a belt around
an inner hexapyridyl template; the porphyrins contain aryl groups
(3,5-di(tri-hexylsilane)benzene) at meso-positions and are interconnected
by acetylenes. The fine structure of the peaks identified as $\S_{1}$
through $\S_{6}$ in its absorption spectrum at 77K (Fig.~\ref{fig:Structure-of-porphyrin}B)
suggests that coherences in 2DES might be clearly resolvable for this
system.

Previous quantum chemistry calculations using the time-dependent density
functional theory estimated the lowest-energy ${\rm S}_{0}-{\rm S}_{1}$
transition to be around $10566\,\icm$. The sequence of almost equally-spaced
strong peaks (for the nanoring studied here found at $11655\,\icm$,
$12295\,\icm$ and $12862\,\icm$) were discussed to stem from the
Franck--Condon progression of the vibrational $\sim605\,\icm$ mode
\cite{Sprafke2011}.

\begin{figure}
\begin{centering}
\includegraphics[width=8cm]{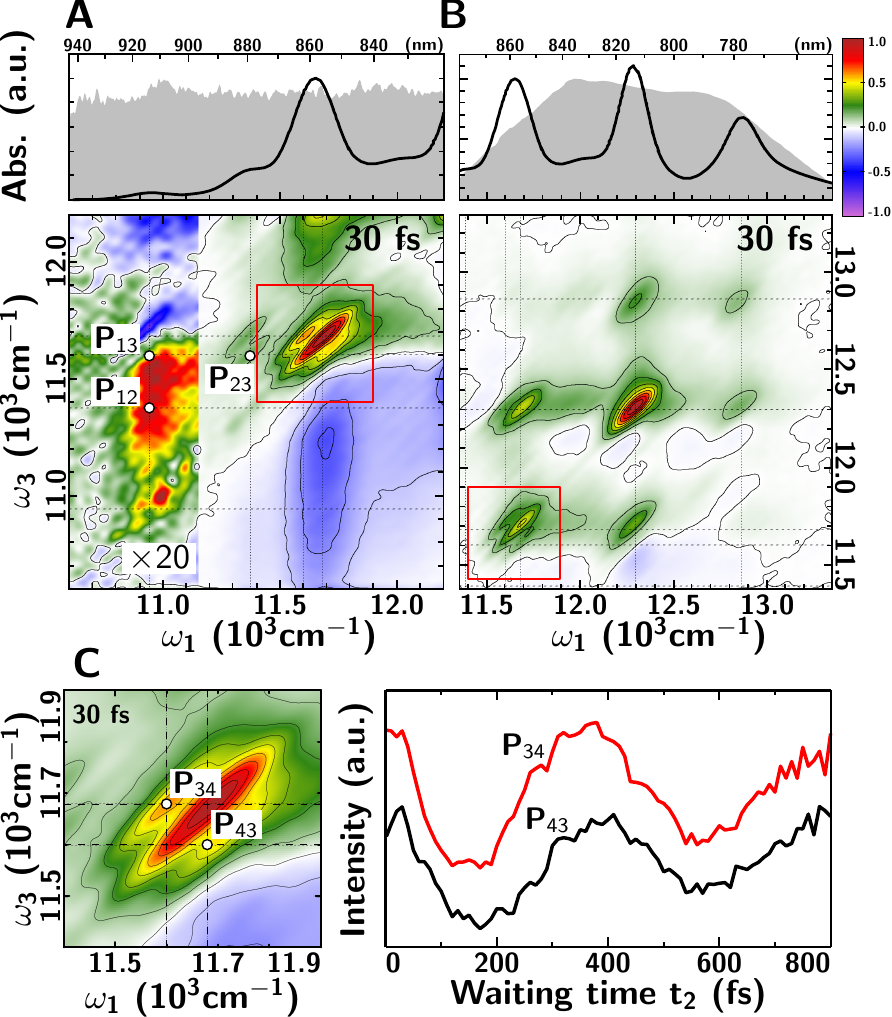}
\par\end{centering}

\protect\caption{\label{fig:2D-spectra-at}\textbf{The absorptive two-dimensional spectra
at different excitation conditions and at 30~fs of the waiting time.}
Spectra were obtained using laser pulses centered at 880~nm (\textbf{A})
and at 800~nm (\textbf{B}), respectively. In the upper panels, laser
pulse and nanoring absorption spectra are shown. In (\textbf{A}) the
signal value is multiplied by a factor of 20 for the plot range where
$\omega_{1}<11157\,\protect\icm$. (\textbf{C}) Zoomed in region of
a degenerate peak in the 800~nm measurement (shown by red squares
in (\textbf{A}) and (\textbf{B})) and oscillatory dynamics of the
``${\rm P}_{34}$'' and ``${\rm P}_{43}$'' peaks. Spectra are
drawn using linear colour scale, normalized to the maximum of each
spectrum. Dashed vertical and horizontal lines indicate the energies
of states $\protect\S_{1}-\protect\S_{6}$.}
\end{figure}

Two separate 2DES measurements of the fully conjugated porphyrin hexamer
nanoring were performed at 77~K using laser pulses, the spectrum
of which was centered either at 800~nm or at 880~nm, thus covering
different parts of the absorption spectrum (refer to the upper panels
in Fig.~\ref{fig:2D-spectra-at}A-B for the corresponding laser spectra).
Refer to the Supplementary Materials and Methods section.

The 2D spectrum obtained using laser pulses at 880~nm covers the
lowest electronic transition (Fig.~\ref{fig:2D-spectra-at}A). It
is dominated by a strong diagonal peak at $\sim11655\,\icm$. Other
features on the diagonal, related to the absorption peaks at $10941\,\icm$
($\S_{1}$) and $11373\,\icm$ ($\S_{2}$), are much weaker. However,
the peaks above the diagonal (``${\rm P}_{12}$'', ``${\rm P}_{13}$''
and ``${\rm P}_{23}$'') connecting the three diagonal peaks can
be clearly resolved in Fig.~\ref{fig:2D-spectra-at}A. Excited state
absorption shows up as strong negative features below the diagonal
overlapping with positive peaks.

Laser pulses with the spectrum centered at 800~nm were used to investigate
the spectral range of the other three most prominent transitions.
The corresponding 2D spectrum is shown in Fig.~\ref{fig:2D-spectra-at}B.
The spectrum is very rich in features and at least 17 peaks can be
clearly resolved. Interestingly, a peak on the diagonal at around
11655~$\icm$ consists of two previously not resolved\cite{Hoffmann2008,Sprafke2011}
contributions separated by $\sim80\,\icm$. It could be estimated
from the position of the off-diagonal peaks, indicated as ${\rm P_{34}}$
and ${\rm P_{43}}$ in Fig.~\ref{fig:2D-spectra-at}C that the energies
corresponding to these states are around 11600~$\icm$ ($\S_{3}$)
and 11680~$\icm$ ($\S_{4}$).

To determine the energy dynamics via the manifold of excited states,
the 2DES data were taken as a function of the waiting time $t_{2}$
, thus, making the complete 3D spectrum. Feature-rich oscillatory
evolution has been observed throughout the whole $(\omega_{1},\omega_{3})$
2D area as a function of $t_{2}$. Decaying dynamics was extracted
using three-exponential-decay fitting with one variable time constant
of 156--250~fs and remaining lifetimes of 235~ps and $\gg$1~ns.
The shortest timescale is related to the downward energy relaxation
in the exciton manifold. This process is observed as the decay of
the peaks on the diagonal of the spectrum and simultaneous increase
of the peaks away from the diagonal if no other competing channel
exists\cite{Brixner2005,Dostal2014}. The longer timescales represent
the relaxation from the lowest state of the exciton manifold to the
ground or the other (for example, triplet) state (for more details
see the Supplementary Text online). 

Coherent beatings in the 2D spectra can be conveniently visualised
by applying Fourier transforms $t_{2}\to\omega_{2}$ to the residuals
\cite{Seibt-JPCC2013,Camargo2015}. The resulting Fourier amplitude
dependency is a three-dimensional $(\omega_{1},\omega_{2},\omega_{3})$
bulk spectrum, the 2D slices ($\omega_{1}$,$\omega_{3}$), or the
so-called oscillation maps, of which can be plotted for a fixed frequency
$\omega_{2}$. Notice that evolution of Hermitian conjugate coherences
$|a\rangle\langle b|$ and $|b\rangle\langle a|$ appears at positive
and negative $\omega_{2}$ frequencies, respectively\cite{Cundiff-3Dspec-ncomms2012}. 

\begin{figure*}
\begin{centering}
\includegraphics[width=16cm]{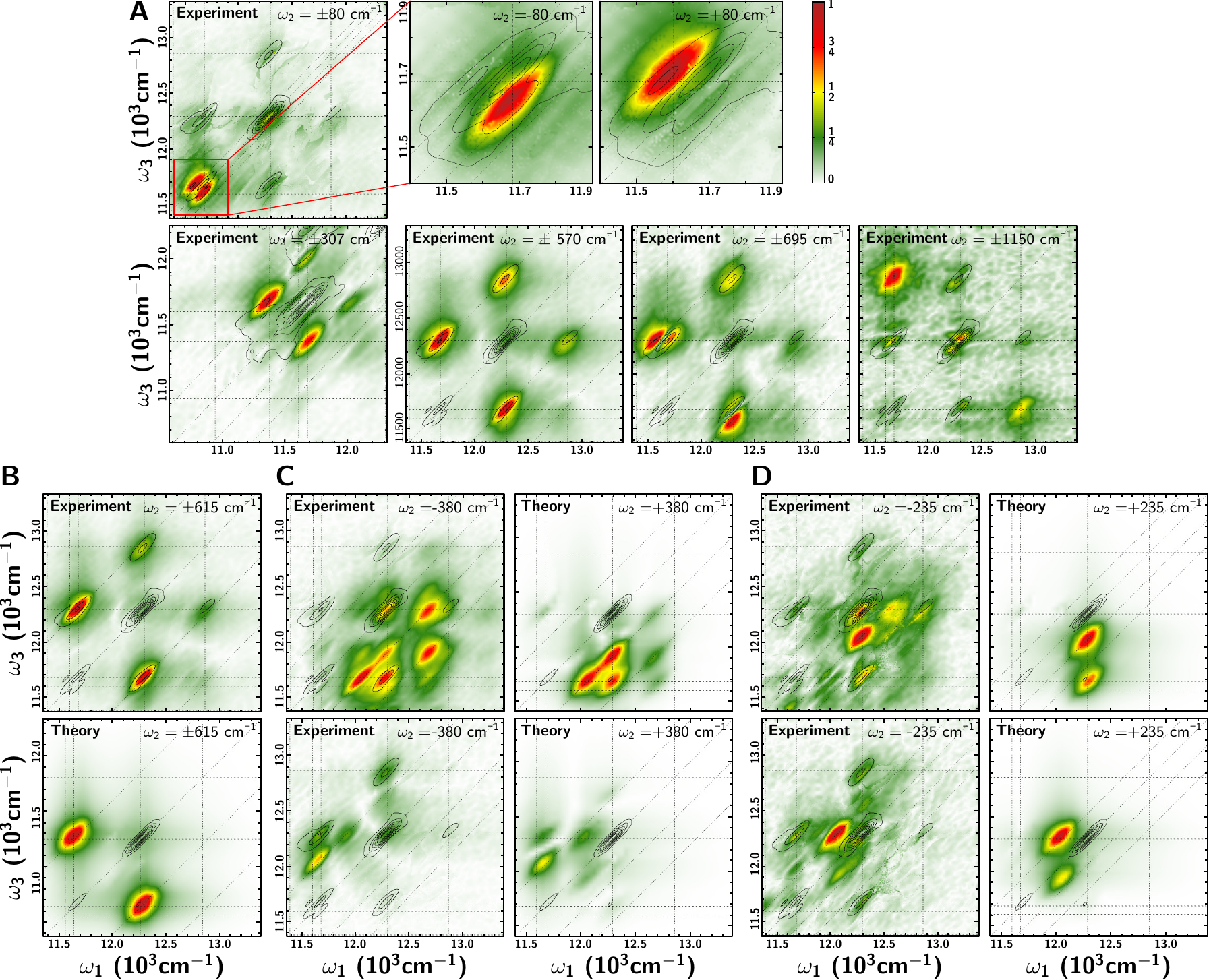}
\par\end{centering}

\protect\caption{\label{fig:maps}\textbf{Diversity of coherences revealed by oscillation
maps with different feature patterns.} \textbf{(A)} experimental coherence
maps of \emph{electronic }coherences at $\omega_{2}=\pm80$, $\pm307$,
$\pm570$, $\pm695$, and $\pm1150\,\protect\icm$; \textbf{(B)} experimental
and simulated oscillation maps for \emph{electronic }coherence at
$\omega_{2}=\pm615\,\protect\icm$; \textbf{(C)} \emph{vibrational}
coherence at $\omega_{2}=\pm380\,\protect\icm$; \textbf{(D)} \emph{mixed
vibronic} coherence at $\omega_{2}=\pm235\,\protect\icm$. Intensity
of oscillations at each point of a map is indicated by the colour
scale; background contours show 2D rephasing spectrum at waiting time
30~fs. Each experimental and simulated map for particular frequency
is independently normalized to the maximal amplitude of either positive
or negative frequency map. Dashed lines parallel to the diagonal are
separated by the value of $\omega_{2}$. Dashed vertical and horizontal
lines indicate the energies of transitions to electronic states $\protect\S_{1}$
through $\protect\S_{6}$.}
\end{figure*}

The 2D slices at a few selected $\omega_{2}$ frequencies are shown
in Fig.~\ref{fig:maps}A-D. Judging by their pattern, three types
of peak configurations can be distinguished. (i) The maps at $\omega_{2}=\pm80$,
$\pm307$, $\pm570$, $\pm615$, $\pm695$, and $\pm1150\,\icm$ shown
in Fig.~\ref{fig:maps}A-B are \emph{diagonally symmetric}, i. e.
positive and negative $\omega_{2}$ features are mirror images of
each other with respect to the diagonal as specifically showed in
Fig.~\ref{fig:maps}A for $\omega_{2}=\pm80\,\icm$. (ii) Oscillation
maps at $\omega_{2}=+380\,\icm$ and $-380\,\icm$, shown in Fig.~\ref{fig:maps}C,
are highly \emph{diagonally asymmetric} with the features below the
diagonal in the $\omega_{2}=-380\,\icm$ map significantly stronger
than in the $\omega_{2}=+380\,\icm$ map. (iii) Features in the $\omega_{2}=-235\,\icm$
and $\omega_{2}=+235\,\icm$ maps (Fig.~\ref{fig:maps}D) are \emph{diagonally
asymmetric}, but their amplitudes are of the similar magnitude for
positive and negative frequency maps. Refer to the Supplementary Fig.~S1
for oscillation maps at all other distinguishable frequencies.

\begin{table}
\raggedright{}\protect\caption{\label{tab:Classification-of-observed-1}\textbf{Classification of
the observed coherences ($<$700~$\protect\icm$).} Average of four
separate measurements are considered and the extracted dephasing times
and standard deviations $\sigma_{\tau}$ are pointed out. The dephasing
times of a few coherences could not be extracted with a reliable experimental
error due to the limited resolution. Dephasing times were obtained
by fitting complex oscillatory signals in the time domain (for details,
see the Supplementary Text online).}
\begin{tabular*}{1\linewidth}{@{\extracolsep{\fill}}cccl}
\hline 
$\omega_{2}$ ($\icm$) & Dephasing time (fs) & $\sigma_{\tau}$ (fs) & Origin of coherence\tabularnewline
\hline 
$\pm$80 & $\gtrsim500$ & -- & Electronic $|\S_{3}\rangle\langle\S_{4}|$\tabularnewline
$\pm$180 & -- & -- & Vibrational\tabularnewline
$\pm$235 & 360 & 160 & Mixed $|\S_{4}^{*}\rangle\langle\S_{5}|$\tabularnewline
$\pm$307 & 280 & 130 & Electronic $|\S_{2}\rangle\langle\S_{4}|$\tabularnewline
$\pm$380 & $>600$ & -- & Vibrational\tabularnewline
$\pm$450 & -- & -- & Vibrational\tabularnewline
$\pm$570 & 80 & 20 & Electronic $|\S_{5}\rangle\langle\S_{6}|$\tabularnewline
$\pm$615 & 200 & 30 & Electronic $|\S_{4}\rangle\langle\S_{5}|$\tabularnewline
$\pm$695 & 190 & 40 & Electronic $|\S_{3}\rangle\langle\S_{5}|$\tabularnewline
\hline 
\end{tabular*}
\end{table}

We have previously shown that electronic, vibrational, and mixed coherences
are, in theory, manifested by their characteristic patterns and symmetries
in the coherence maps\cite{Butkus_vibronic_aggregatess_JCP2014}.
The symmetry of the experimental maps at $\omega_{2}=\pm80$, $\pm307$,
$\pm570$, $\pm615$, $\pm695\,\icm$, and $\pm1150\,\icm$ (Fig.~\ref{fig:maps}A
and B) indicate that the underlying coherences are of the electronic
origin. Dephasing times of these coherences are shorter than 300~fs
(Table~\ref{tab:Classification-of-observed-1}), oscillations are
present only in the off-diagonal regions in the rephasing 2D spectrum
(and only in the diagonal peaks in the non-rephasing 2D spectrum;
see Supplementary Fig.~S2), and the peaks at positive and negative
$\omega_{2}$ frequencies are of the same amplitude.

The electronic coherence with $\omega_{2}=+80\,\icm$ (and $\omega_{2}=-80\,\icm$)
can be assigned to the coherent superposition $|\S_{4}\rangle\langle\S_{3}|$
(and $|\S_{3}\rangle\langle\S_{4}|$) of the closely-positioned electronic
states in the vicinity of $11655\,\icm$. The electronic coherence
at $\omega_{2}=\pm307\,\icm$ shows up only in the measurement using
laser pulses at 880~nm, signifying electronic quantum beats between
states $|\S_{2}\rangle$ and $|\S_{4}\rangle$, that could not be
excited by laser pulses at 800~nm. 

Beatings with the $\pm570$, $\pm615$, $\pm695\,\icm$ and $\pm1150\,\icm$
frequencies represent quantum coherences $|\S_{5}\rangle\langle\S_{6}|$,
$|\S_{4}\rangle\langle\S_{5}|$, $|\S_{3}\rangle\langle\S_{5}|$ and
$|\S_{4}\rangle\langle\S_{6}|$ (and their Hermitian conjugates).
We find crosstalk in oscillation maps at $\omega_{2}=\pm570$, $\pm615$,
and $\pm695\,\icm$ due to the overlapping of different beating frequencies
and limited frequency resolution resulting from the damped dynamic
coherences. For example, two weaker peaks in the map of $\omega_{2}=\pm615\,\icm$
indicate the crosstalk from coherences $|\S_{4}\rangle\langle\S_{5}|$
and $|\S_{5}\rangle\langle\S_{4}|$ with $\pm570\,\icm$ frequency
(Fig.~\ref{fig:maps}B). The crosstalk between different frequencies
is supported by observation that the weaker peaks are shifted away
from the $\omega_{3}=\omega_{1}\pm615\,\icm$ dashed lines, parallel
to the diagonal. Thus, our findings imply that states $|\S_{1}\rangle$
through $|\S_{6}\rangle$ are of electronic origin in contrast to
previous assignment\cite{Hoffmann2008,Sprafke2011} (it should be
noted that the previously suggested vibrational progression does not
follow a displaced harmonic oscillator model). 

We assign beatings with the $\omega_{2}=\pm380\,\icm$ frequency to
the vibrational coherence. This follows from the oscillation maps
(Fig.~\ref{fig:maps}C), which have the pattern of oscillating peaks
typical for the vibrational coherence\cite{Butkus-Zigmantas-Abramavicius-Valkunas-CPL2012}:
the $\omega_{2}=-380\,\icm$ map contains many features below the
diagonal, while the $\omega_{2}=+380\,\icm$ map is similar to the
electronic coherence maps presented in Fig.~\ref{fig:maps}A and
B. In contrast to the electronic coherences, the amplitude of the
$\omega_{2}=-380\,\icm$ map is significantly stronger that that of
$\omega_{2}=+380\,\icm$ \cite{Jonas_PNAS2012,Butkus2013}. Detailed
analysis of the map implies that the vibrational ground state coherences
$|{\rm g}\rangle\langle{\rm g}^{*}|$ ($|{\rm g}^{*}\rangle$ denotes
some vibrationally hot ground state) appear exclusively at $\omega_{2}<0$.
Their strong amplitudes are therefore related to the long lifetimes
of the ground state vibrations compared with the ones at electronic
excited states and mapped onto $\omega_{2}=+380\,\icm$. Similar maps
at $\omega_{2}=$ $180\,\icm$, $450\,\icm$, and $835\,\icm$ (see
the Supplementary Text online) imply their vibrational origin as well.

However, beatings at $\omega_{2}=\pm235\,\icm$ point out to the mixed
coherence, signifying the superposition state of the $|\S_{5}\rangle$
electronic state and vibronically hot state $|\S_{4}^{*}\rangle$
($380\,\icm$ vibrational mode). The corresponding oscillation map
is not typical of neither vibrational nor electronic coherences (see
Fig.~\ref{fig:maps}D) and the beating frequency $235\,\icm$ is
equal to the difference between the corresponding states.

To support the assignment of the excited states and the natures of
the corresponding quantum coherences, two different theoretical models
were considered: the \emph{electronic-only} model of six excitonically
coupled porphyrin molecules and the \emph{vibronically coupled} dimer
model. The electronic-only model is used to capture the connection
between the cyclic symmetry of the system and its optical response.
The vibronic model supports assignment of the vibrational, electronic,
and mixed features. 

In the electronic-only model, the nanoring is represented by six two-level
sites, representing porphyrins in a circular arrangement, each site
characterized by a single transition dipole, rotated by an angle $\alpha=49^{\circ}$
with respect to the tangent of the nanoring backbone. Two types of
deformations---twisting and bending of the ring backbone---were taken
into consideration (refer to the inset in Fig.~\ref{fig:Structure-of-porphyrin}A).
Due to the full $\pi$-conjugation of the nanoring, the interaction
between any two porphyrins cannot be described by the dipole--dipole
approximation and appropriate scaling factors of the coupling constants
calculated in the dipole--dipole approximation for the nearest, next-nearest,
and next-next-nearest neighbors were obtained by fitting the electronic-only
model to the experimental absorption spectrum (see the Supplementary
Text online for more details). The simulated absorption spectrum of
this model is presented in Fig.~\ref{fig:Structure-of-porphyrin}B
by the red dashed line. The agreement with the experimental absorption
spectrum is quite good except of the intensity of the $\S_{5}$ transition.
The absence of vibronic coupling in the electronic-only model may
be responsible for this discrepancy. Particularly, the electronic
model provides the basis of the observed electronic coherences described
above. However, additional vibrational/vibronic ingredients are necessary
to explain the remaining coherences.

For qualitative simulations of the vibronic coupling in the porphyrin
nanoring we used the theoretical approach developed for a vibronic
molecular dimer\cite{Butkus_vibronic_aggregatess_JCP2014,Basinskaite2014},
extended by including two vibrational modes. We assume that two electronic
states with the highest oscillator strengths, $\S_{4}$ and $\S_{5}$,
are coupled to two vibrational modes of $380\,\icm$ and $450\,\icm$
with the Huang--Rhys factors of 0.03 and 0.01, respectively. This
allows to significantly improve the description of the absorption
spectrum in the range of $\S_{4}$ and $\S_{5}$ peaks (Fig.~\ref{fig:Structure-of-porphyrin}B).
We also performed simulations of the 3D spectra and extracted its
2D slices at various $\omega_{2}$ values, corresponding to electronic,
vibrational and mixed coherences (Fig.~\ref{fig:maps}B, C and D).
Although we included only two electronic states in the vibronic model,
calculated and experimental maps of the vibrational coherences at
$\pm380\,\icm$, the electronic coherences at $\pm615\,\icm$, and
the mixed coherences at $\pm235\,\icm$ are in a very good agreement
with the experimental ones, confirming our assignments.

By this analysis we prove the coexistence of electronic, vibrational
and mixed coherences in a single system of the porphyrin hexamer.
Analysis of quantum coherences aids in disentangling the energy level
structure of the excited states and their cooperativity. This turns
out to manifest even in the absorption spectrum, where we identify
electronic transitions, the intensity and positions of which are non-trivially
defined by vibronic coupling (together with small ring deformations).
In the 3D measurement the whole ``zoo'' of coherences gets raised
and they inter-operate to maintain long coherence lifetimes and coherent
excitation evolution. Such coherent quantum properties of a supermolecular
system are reported for the first time, but should be general for
molecular aggregates.

\section*{Acknowledgments}

D.~Z. and J.~A. were supported by Swedish Research Council, the
Knut and Alice Wallenberg Foundation, the Wenner--Gren Foundation,
and partially funded by the European Social Fund under the Global
grant measure. The work in Oxford was supported by European Commission
(Marie Curie Individual Fellowship to P.~N. under contract PIEF-GA-2011-301336).
V.~B. and D.~A. acknowledge the support of the European Social Fund
under the Global Grant Measure (No. VP1-3.1-\v{S}MM-07-K-01-020).

\end{document}